\documentclass[aps,prl,reprint,amsmath,amssymb,
superscriptaddress,showpacs,showkeys]{revtex4-2}
\usepackage{graphicx}
\usepackage{dcolumn}
\usepackage{bm}
\usepackage{soul} 
\usepackage{color}
\definecolor{vs}{rgb}{0.1,0.4,0.1}                  

\newcommand{\del}[1]{}                             
\usepackage{hyperref}
\usepackage{verbatim}
\newcommand\wordcount{\verbatiminput{\jobname.sum}}
\begin{document}


\title{Fall of Quantum Particle to the Center: Exact solution}


\author{Michael I. Tribelsky}
\email[Corresponding author (replace ``\_at\_" by @):\\]{\mbox{E-mail: mitribel\_at\_gmail.com}
}
\homepage[]{https://polly.phys.msu.ru/en/labs/Tribelsky/}
\affiliation{
M. V. Lomonosov Moscow State University, Moscow, 119991, Russia}
\affiliation{National Research Nuclear University MEPhI (Moscow Engineering Physics Institute), Moscow, 115409, Russia}

\date{\today}

\begin{abstract}A fall of a particle to the center of a singular potential is one of a few fundamental problems of quantum mechanics. Nonetheless, its solution is not complete yet. The known results just indicate that if the singularity of the potential is strong enough, the spectrum of the Schrodinger equation is not bounded from below. However, the wave functions of the problem do not admit the limiting transition to the ground state. Therefore, the unboundedness of the spectrum is only a necessary condition. To prove that a quantum particle indeed can fall to the center, a wave function describing the fall should be obtained explicitly. This is done in the present paper. Specifically, an exact solution of the time-dependent Schrodinger equation corresponding to the fall is obtained and analyzed. A law for the collapse of the region of the wave function localization to a single point is obtained explicitly. It is shown that the known necessary conditions for the particle to fall simultaneously are sufficient.
\end{abstract}


\maketitle
%
%
{\it Introduction.} It is impossible to overrate the importance of the Schr\"{o}dinger equation (SE). To say nothing that it is the keystone of entire non-relativistic quantum mechanics, it arises in many other quite classical problems, for example, in the stability analysis of solutions of various nonlinear problems in combustion~\cite{zeldovich1959theorv}, optical damage of glass initiated by small inclusions~\cite{anisimov1973role}, Gunn domains in semiconductors~\cite{knight1967theory,Tribelsky2021}, structure stability of the interphase boundary in liquid crystals~\cite{kats1990structure}, etc. Regarding the analogy between the quantum scattering of a particle and the light scattering by an obstacle, this is a well-known fact.

Among the vast diversity of problems related to SE and its applications, there are several revealing its fundamental properties. They have paramount importance and enter into all main textbooks on quantum mechanics. The problem of a ``fall" of a particle to the origin of a singular potential, also known as {\it quantum collapse}, is the one of such a kind, see, e.g., Ref.~\cite{landau2013quantum}. Nowadays, in this problem, the interest is shifted to its various generalizations 
~\cite{PhysRevLett.85.1590,sakaguchi2011suppression,sakaguchi2011suppression2,sakaguchi2013suppression3,astrakharchik2015quantum,Malomed2018,Shamriz2020,shamriz2020suppression}. 
However, the critical question of whether the collapse indeed may happen actually remains open even in the simplest case of a single quantum particle in a centrally symmetric potential; see below. The goal of this paper is to fill the gap by obtaining an exact solution to this problem.

{\it Preliminary.} In classical mechanics, the problem of a fall to the origin of a particle moving in a centrally symmetric potential $U(r)$ is trivial since the governing equations are exactly integrable at any $U(r)$~\cite{landau1976mechanics}. The integration shows that the particle may pass through the origin, provided
\begin{equation}\label{eq:class}
  [r^2U(r)]_{r\rightarrow 0} <-{M^2}/(2m),
\end{equation}
where $M$ is the angular momentum and $m$ stands for the particle mass~\cite{landau1976mechanics}.

In quantum mechanics, the case is more tricky. Firstly, just a few potentials make SE exactly integrable. Secondly, a particle's localization in a given space point must not contradict the uncertainty relations. The latter implies a divergence of the uncertainty of the particle momentum if the localization occurs.

Estimates of the spectrum of SE equation with
\begin{equation}\label{eq:U}
  U(r) = -\beta/r^s;\;\; (\beta > 0)
\end{equation}
reveal that at $s>2$, there are bound states with energy $E<0$, which is not limited from below. This indicates a {\it possibility\/} of the collapse. At $s<2$, the spectrum is bounded from below, and the collapse cannot happen. The case $s=2$ requires more accurate consideration~\cite{landau2013quantum}.

It should be stressed that potential \eqref{eq:U} with \mbox{$s\geq 2$} does exist in nature. For example, the potential singular as $1/r^2$  describes the interaction of a point electric charge with a particle with zero total electric charge but a fixed finite dipole moment~\cite{sakaguchi2011suppression}, to say nothing about various applications of SE in non-quantum problems. Thus, in addition to purely academic interest, the quantum collapse may have a pretty practical meaning.

Note now that while the spectrum bounded from below does mean that the fall to the center is not possible, the unboundedness of the spectrum from below is just {\it the necessary condition\/} for the fall to happen. To make the fall sure, one has to build the corresponding wave function explicitly. The latter is hardly possible with the help of the stationary-state eigenfunctions since for the given problem they do not have any limit at $r \rightarrow 0$. This behavior prevents one from implementing a direct limiting transition to the ground state with $E \rightarrow -\infty$.

On top of that, even if the expression for the stationary-state eigenfunction with $E = -\infty$ were obtained, it would not mean yet that the fall can indeed occur. To prove the occurrence of the fall, one has to show that there is at least a single initial state whose wave function has a non-zero projection to this eigenfunction, i.e., that the state corresponding to the localization of the particle at $r=0$ may be indeed {\it excited}. Last but not least is the question about the law describing the contraction with the time of the region of the wave function spatial localization if the fall does happen. Thus, strictly speaking, the question about a quantum particle fall to the origin of a centrally symmetric potential remains open.

The difficulties mentioned above have arisen owing to the attempt to describe an essentially time-dependent process of the fall with the help of the stationary-state eigenfunctions. In a sense, it looks like trying to force a square peg into a round hole. To save the day, one has to consider a spatiotemporal evolution of a wave package, and this consideration should not be based on the projection of the corresponding wave function onto the stationary-state eigenmodes.

Here I should confess that the idea is not mine. When I was a junior scientist, my adviser Dr. Sergei I. Anisimov~\cite{Dzyaloshinskii_2019} from Landau Institute, told me about this approach to the problem. Moreover, he said that together with Dr. Igor E. Dzyaloshinskii~\cite{Dzyal_2021}, they had found a solution describing the collapse of the wave package.

Many years later, in connection with light scattering by nanoparticles, I came across the problem mathematically analogous to the quantum collapse. I remembered this conversation with Dr. Anisimov and asked him for details and references. The reply was that these results had never been published and details he did not remember. Now, when both Dr. Anisimov and Dr. Dzyalashinskii passed away, I decide to reobtain their results and make them available to a broad readership as a small token of my great respect to these distinguished scholars.

{\it Problem formulation.} Thus, we have to start from the full time-dependent version of SE:
\begin{equation}\label{eq:Schr}
 i\hbar \frac{\partial \Psi}{\partial t} = \hat{H}\Psi;\;\; \hat{H} \equiv -\frac{\hbar^2}{2m}\Delta + U(r),
\end{equation}
where $\Delta$ stands for the Laplasian. Eq.~\eqref{eq:Schr} should be supplemented by the initial condition $\Psi(\mathbf{r},t) = \Psi_0(\mathbf{r})$ and the standard boundary conditions stipulating convergence of $\int |\Psi|^2d^3r$ for the states corresponding to the bounded in space motion.

Suppose that the fall does take place. It means the collapse of the region of the wave function localization to a point. Then, in the course of the collapse, the wave function ``forgets" the characteristic spatial scale of $\Psi_0(\mathbf{r})$, if any. In other words, at the final stage of the collapse, the dynamic of the wave function may become {\it self-similar\/} and hence {\it universal}, i.e., independent of the initial condition, provided the latter gives rise to the collapse.

Then, we have to check whether Eq.~\eqref{eq:Schr} has self-similar solutions and, if so, to find them explicitly. To prove the possibility of the fall, obtaining just a single solution of such a kind is enough.

{\it Self-similar problem}. We count the time down from the moment of the complete collapse so that the collapse corresponds to $t=0$, and its dynamics is described by $-\infty <t<0$. Once again, we consider the potential \eqref{eq:U}. Let us look for a self-similar solution in the form
\begin{equation}\label{eq:Self_Psi}
  \Psi = R(\xi)Y_\ell^m(\theta,\varphi).
\end{equation}
Here $Y_\ell^m(\theta,\varphi)$ is the spherical harmonic function and $\xi = r/(-\chi t)^\nu$, where $\chi$ and $\nu$ are unknown yet constants. Substitution of Eq.~\eqref{eq:Self_Psi} into Eq.~\eqref{eq:Schr} indicates that the resulting equation for $R(\xi)$ is reduced to an ODE depending on the single self-similar variable $\xi$ solely if and only if
\begin{equation}\label{eq:param}
  s=2,\;\nu = 1/2,\; \chi = const \cdot \hbar/m.
\end{equation}
The dimensionless constant in the expression for $\chi$ may be any. In what follows, it is convenient to suppose it equals unity. Then, the corresponding governing equation for $R(\xi)$ reads:
\begin{equation}\label{eq:R}
  R''+\left(\frac{2}{\xi}+i\xi\right)R'+\frac{\gamma R}{\xi^2}=0,
\end{equation}
where $\gamma \equiv \beta-\ell(\ell+1)$ and prime denotes $d/d\xi$.
\begin{figure*}
  \centering
  \includegraphics[width=\textwidth]{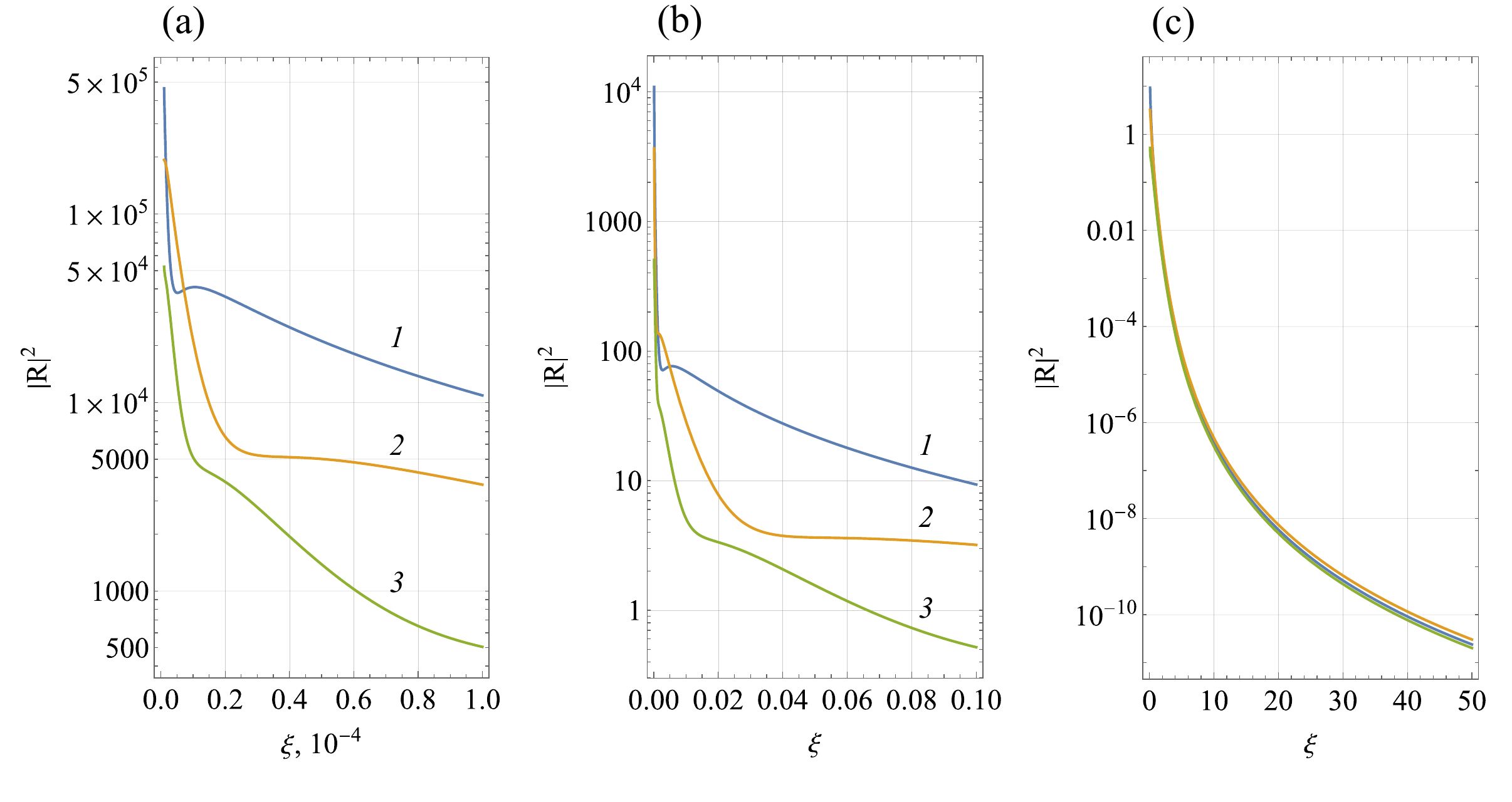}
  \caption{$R(\xi)$ in several characteristic ranges of $\xi$; \mbox{$\gamma = 1/2$~(1),} $\gamma = 1$~(2), and $\gamma = 2$~(3).}\label{fid:R}
\end{figure*}

Eq.~\eqref{eq:Self_Psi} is exactly integrable. Its solution satisfying the condition $R(\xi)\rightarrow 0$ at $\xi \rightarrow \infty$ is
\begin{widetext}
\begin{eqnarray}
R(\xi) & = &\frac{C}{\sqrt{\xi}} \left[{\xi}^{-\frac{i \alpha }{2}}{2}^{\frac{i \alpha }{2}}\Gamma \left(1+\frac{ i \alpha }{2} \right) \Gamma \left(\frac{5-i \alpha}{4}\right)\! \, _1F_1\left(-\frac{1+i \alpha
   }{4};\frac{2-i \alpha }{2};-\frac{i \xi^2}{2}\right)\right. \label{eq:Sol_R} \\
   & - &\left.\xi^{\frac{i\alpha}{2} }\exp\left(-\frac{\pi\alpha}{4}\right)
   \Gamma \left(1-\frac{i \alpha }{2}\right) \Gamma \left(\frac{5+i \alpha}{4}\right) \! \, _1F_1\left(-\frac{1-i \alpha}{4} ;\frac{2+i \alpha }{2};-\frac{i
   \xi^2}{2}\right)\right]. \nonumber
\end{eqnarray}
\end{widetext}
Here $C$ is a constant of integration, $\Gamma(z)$ stands for the Euler gamma function, $_1F_1(a,b,z)$ designate  the Kummer confluent hypergeometric function of the first kind~\cite{Buchholz1969}, and $\alpha \equiv \sqrt{4\gamma - 1}>0$. The positiveness of $\alpha$ follows from the fact that for $U(r) = \beta/r^2$ the necessary conditions for the fall to occur derived from the unboundedness of the spectrum from below stipulate that $\gamma > 1/4$~\cite{landau2013quantum}.

Eq.~\eqref{eq:Sol_R} requires a comment: It includes real positive quantities 2 and $\xi$ in imaginary power $\pm i\alpha/2$. These expressions may be transformed as follows:
\begin{eqnarray}
  z^{\pm i\frac{\alpha}{2}}&=&(e^{i2\pi n+\log z})^{\pm i\frac{\alpha}{2}} = e^{\mp\pi\alpha n}e^{i\frac{\alpha}{2}\log z} \label{eq:z^ia} \\
  &=&e^{\mp\pi\alpha n}\left[\cos\left( \frac{\alpha}{2}\log z\right) \mp i\sin\left(\frac{\alpha}{2}\log z\right)\right],\nonumber
\end{eqnarray}
where $n$ is arbitrary integer and $z>0$.

Eq.~\eqref{eq:z^ia} has an infinite number of branches corresponding to different values of $n$. At any $n$, all these branches have the same phase equal to $({\alpha}/{2})\log z$. Therefore, the real and imaginary parts of any branch have a number of zeros which in the same manner increases without limit as $z \rightarrow 0$. However, the modulus of these branches varies for different $n$ as $\exp({\mp\pi\alpha n})$.

For the sake of simplicity, anywhere in Eq.~\eqref{eq:Sol_R} and in what follows, only the single branch with $n=0$ is selected. Just for this branch Eq.~\eqref{eq:Sol_R} satisfies the condition  $R(\xi)\rightarrow 0$ at $\xi \rightarrow \infty$. For other branches to fulfill this condition, the prefactor at the Kummer functions in Eq.~\eqref{eq:Sol_R} should be changed accordingly.

Eq.~\eqref{eq:Sol_R} is a solution to the problem, provided the obtained wave function may be normalized, i.e., $\int_{0}^{\infty} |R(\xi)|^2\xi^2d\xi$ converges. 
Based on the explicit form of $R(\xi)$ given by Eq.~\eqref{eq:Sol_R} one may select just two ``dangerous" areas, which might result in divergence of the integral, namely the asymptotics at $\xi \rightarrow \infty$ and at $\xi \rightarrow 0$.

Analysis of the behavior of the integrand at $\xi \rightarrow \infty$ is relatively trivial. Employing the known asymptotical expression for $_1F_1(a,b,z)$~\cite{Buchholz1969} one readily obtains that
\begin{equation}\label{eq:R_infin}
  R(\xi) = C_\infty \frac{\exp(i\xi^2/2)}{\xi^3} + \ldots,
\end{equation}
where $C_\infty$ is a constant and ellipsis indicates dropped higher order in $1/\xi$ terms. Thus, at the upper limit $\int_{0}^{\infty} |R(\xi)|^2\xi^2d\xi$ converges.

The case $\xi \rightarrow 0$ is more tricky owing to the singularity described by Eq.~\eqref{eq:z^ia} . The asymptotical behavior of Eq.~\eqref{eq:R} at $\xi \rightarrow 0$ is

\begin{eqnarray}
  R(\xi) &=& \frac{C}{\sqrt{\xi}} \left[{\xi}^{-\frac{i \alpha }{2}}{2}^{\frac{i \alpha }{2}}\Gamma \left(1+\frac{ i \alpha }{2} \right) \Gamma \left(\frac{5-i \alpha}{4}\right)\right. \label{eq:R0}\\
   &-& \left.\xi^{\frac{i\alpha}{2}}\exp\left(-\frac{\pi\alpha}{4}\right)\Gamma \left(1-\frac{i \alpha }{2}\right) \Gamma \left(\frac{5+i \alpha}{4}\right)\right].\nonumber
\end{eqnarray}

Importantly, however, that according to Eqs.~\eqref{eq:z^ia}, \eqref{eq:R0} $|R(\xi)|^2$ in the vicinity of $\xi = 0$ is majorized by $C_0/\xi$, where $C_0$ is a constant. Therefore, the singularity at $\xi = 0$ is integrable, and $\int_{0}^{\infty} |R(\xi)|^2\xi^2d\xi$ converges at the lower limit too. In other words, at any finite $t<0$ the quantity $\langle \Psi|\Psi\rangle$ remains finite.

As an example, plots $|R(\xi)|^2$ at several values of $\gamma$ are presented in Fig.~\ref{fid:R}. It is interesting to note that despite Re$\,R(\xi)$ and Im$\,R(\xi)$ both exhibit an infinite number of oscillations as $\xi$ tends to zero, this is not the case for $|R(\xi)|^2$, which is a monotonically decaying function of $\xi$.

{\it Discussion.} The obtained $\Psi$ makes it possible to calculate the average value of any operator $\hat{A}$: \mbox{$\langle \hat{A} \rangle =  \langle \Psi| \hat{A} |\Psi \rangle/\langle \Psi|\Psi \rangle$}, provided the integral $\langle \Psi| \hat{A} |\Psi \rangle$ converges, i.e., $\langle \hat{A} \rangle$ exists. However, it should be remembered that $\Psi =  R(\xi)Y_\ell^m(\theta,\varphi)$, while in scalar products the integration is taken over $\mathbf{r}$-space, which requires the corresponding replacement of the variable. Because of that the scalar products become time-dependent. Then, one readily obtains that
\begin{equation}\label{eq:averages}
  \langle \hat{r} \rangle =C_r\sqrt{-\chi t},\;\;\langle \hat{p} \rangle =\hbar C_p/\sqrt{-\chi t},
\end{equation}
(remember that $t<0$). Here $\hat{p}$ stands for modulus of the momentum and $C_{r,p}$ are dimensionless constants of the order of unity. Thus, the size of the spatial region of the particle localization contracts as $\sqrt{-t}$ while its momentum diverges as $1/\sqrt{-t}$ so that $\Delta p \Delta r$ remains a constant of the order of $\hbar$, in agreement with the uncertainty relations.

Lets us now calculate the average energy of the falling particle~\footnote{The discussed quantum state is not stationary; its energy is not an eigenvalue of the Hamiltonian. Therefore for this state only average energy is meaningful.}, namely $E=\langle \hat{H} \rangle$. According to Eq.~\eqref{eq:Schr}
\begin{equation}\label{E_Schr}
 E= \langle \Psi|\hat{H}|\Psi\rangle/\langle \Psi|\Psi\rangle = i\hbar\langle \Psi|\partial\Psi/\partial t\rangle/\langle \Psi|\Psi\rangle.
\end{equation}
Bearing in mind that for the self-similar $\Psi$, given by Eq.~\eqref{eq:Self_Psi}
\begin{equation*}\label{eq:d/dt}
   \frac{\partial}{\partial t} = \frac{\partial}{\partial\xi}\frac{\partial \xi}{\partial t}=-\frac{\xi}{2t}\frac{\partial}{\partial\xi}
\end{equation*}
and employing the asymptotics \eqref{eq:R_infin}, \eqref{eq:R0}, it is easy to see that at any $t<0$ integral $\langle \Psi|\partial\Psi/\partial t\rangle$ converges. Seemingly, this gives rise to a paradoxical conclusion that $E$ is not conserved and varies in time as $1/t$.

However, this is not the case. The point is that for the given quantum state $E=0$. Owing to the very complicated structure of Eq.~\eqref{eq:R} it is rather difficult to prove the vanishing of $E$ by the direct calculation of the corresponding integrals. Much more simple proof may be done based on the general properties of SE. Indeed, suppose in a certain arbitrary moment of time $E$ has a finite negative value. Since $U(r)\rightarrow 0$ at $r \rightarrow \infty$ it means that at $r > r_0$, where $r_0$ satisfies the condition $U(r_0) = E$, the classically inaccessible region lies. In this case, at  $r \gg r_0$, the wave function must exponentially decay. However, Eq.~\eqref{eq:R_infin} does not exhibit an exponentially decaying asymptotic at $\xi \rightarrow \infty$. Similarly, if $E$ is a finite positive quantity, the asymptotic at $\xi \rightarrow \infty$ should correspond to a traveling plane wave, which also disagrees with Eq.~\eqref{eq:R_infin}. Thus, the only remaining option is $E=0$.

{\it Conclusions.} The performed study gives rise to the following outcome. The fall of a quantum particle to the center does take place. In the potential $U(r)= -\beta/r^2$, it occurs under the necessary conditions formulated in Ref.~\cite{landau2013quantum}. It means that these conditions simultaneously are sufficient. The fall is described by the self-similar solution of SE given by Eqs. \eqref{eq:Self_Psi}, \eqref{eq:Sol_R}. In this case, the radius of the wave function localization region collapses as $\sqrt{-\chi t}$, while the characteristic value of the momentum increases as $1/\sqrt{-\chi t}$. Here $t<0$ is the time counted down from the moment of the full collapse.
Finally, note that since SE is invariant for time reversal accompanied with complex conjugation, these transformations applied to \mbox{Eqs.~\eqref{eq:Self_Psi}, \eqref{eq:Sol_R}} generate the wave function describing the escape of the particle from the center at $t>0$.

\begin{acknowledgments}
{\it Acknowledgements}. This work is supported by the Russian Foundation for Basic Research (Projects No. 20-02-00086) for the analytical study, the Russian Science Foundation (Project No. 21-12-00151) for the symbolic computer calculations, the Moscow Engineering Physics Institute Academic Excellence Project (agreement with the Ministry of Education and Science of the Russian Federation of 27 August 2013, Project No. 02.a03.21.0005) for the package of computer graphic and the Russian Science Foundation (Project No. 19-72-30012) for the provision of user facilities.
\end{acknowledgments}
%

\end{document}